\documentstyle[prl,aps,preprint]{revtex}
\begin{document}
\draft

\title{Quasi-Normal Mode Expansion for Linearized Waves
in Gravitational Systems}
\author{E.S.C. Ching${}^{(1)}$, P.T. Leung${}^{(1)}$,  W.M. Suen${}^{(2)}$ and
K. Young${}^{(1)}$}
\address{${}^{(1)}$Department of Physics, The Chinese University of Hong Kong,
Hong Kong}
\address{${}^{(2)}$Department of Physics, Washington University, St Louis, MO
63130, U S A}
\date{\today}
\maketitle

\begin{abstract}

The quasinormal modes (QNM's) of gravitational systems modeled by
the Klein-Gordon equation with effective potentials are studied
in analogy to the QNM's of optical cavities.
Conditions are given for the QNM's to form a complete
set, i.e., for the Green's function to be expressible as
a sum over QNM's, answering a conjecture by Price and Husain [Phys.
Rev. Lett. {\bf 68}, 1973 (1992)].
In the cases where the QNM sum is divergent, procedures for
regularization are given.
The crucial condition for completeness is the existence of
spatial discontinuities
in the system, e.g., the discontinuity at the stellar surface in the
model of Price and Husain.

\end{abstract}



Radiation from optical cavities is often analysed in terms of
the ``modes" of the cavity \cite {opticqnm}.
Because the waves escape, causing the total energy in the cavity to
decrease, these ``modes" are quasi-normal modes (QNM's). The observation of
the QNM's from the outside immediately gives information on the spatial
structure of the cavity (but not the spatial structure of the source of
radiation inside the cavity), e.g., in obvious notation,
$ \omega _{j} \sim j \pi c / L$, where $L$
is the length of a simple 1-dimensional optical cavity.

Similarly, gravitational radiations from relativistic systems can be
described by QNM's.  Although
``cavities" in gravitational systems, i.e., regions of spacetimes which
reflect and scatter waves significantly, tend to be more leaky than
optical cavities, we argued that QNM analysis and the optical analogy
can still be very useful.
Indeed the QNM's of black holes \cite{bhqnm} and relativistic stars
\cite{detweiler}
have been subjects of much study.  In numerical simulations, it is often
found \cite{numerwave1,numerwave2} that the radiation observed in many black
hole processes
is dominated by the QNM's of the hole.  For processes as violent and
non-linear as the head-on collision of two black holes, the domination
of the radiation by QNM's may seem surprising \cite{numerwave2}.
However, it is easy to understand with the optical analogy
--- the distant observer sees only the QNM's of the optical
cavity, but {\it not} the details of the radiation generation mechanism.

Gravitational QNM radiation may be observed by LIGO and VIRGO \cite{nature}
in the next decade, and may therefore reveal spacetime structures of various
gravitational systems, e.g.,  the strong field region around a black hole
--- in much the same way as the spectrum of a laser permits a distant
observer to infer some characteristics of the cavity.

This exciting possibility calls for the study of the general properties
of QNM's of gravitational systems.
For the interpretation and extraction of observational data,
one would like to know, for example, how the QNM frequencies
of a black hole are perturbed by a massive accretion disk around it.

A fundamental question, raised by  Price and Husain \cite{price},
is whether QNM's can form complete sets. This question is not only of
theoretical interest, but also of technical importance, e.g., the application
of the usual Rayleigh perturbation theory to calculate frequency shifts
hinges on the existence of a complete set.
The answer to this question is thought to be in general negative, as
the system giving rise to QNM's is nonhermitian.
Yet Price and Husain \cite{price} have given a model
of relativistic stellar oscillations
that does have a complete set of QNM's.

In this Letter, we consider this question in a general context, and
give conditions for completeness to hold.
The crucial condition is the existence of spatial
discontinuities.  (In the model of \cite{price}, this discontinuity
corresponds to the stellar surface\cite{price2}.)
Such a condition
is not surprising:  a discrete sum of QNM's can at best be
complete over a finite interval; the discontinuities
provide a natural demarcation of this interval, analogous to the
boundaries of an optical cavity.

The propagation of waves in curved space is often modeled by
the Klein-Gordon (KG) equation \cite{kg}

\begin{equation}
D \phi(x,t) \equiv \left[ {\partial^2 _t}
      -{\partial^2 _x}
      +V(x)
\right] \phi (x,t) = 0   ~~~~ ,
\end{equation}

\noindent
with the outgoing wave boundary condition at infinity.
The potential $V(x)$, assumed to be positive, bounded and vanishing at
infinity, describes the scattering
of wave by the background geometry.  We show
that the QNM's of (1) form a complete set, in the sense that the
Green's function can be expressed as a sum over QNM's (cf. (10) below),
provided the following three conditions hold:
(i) $V(x)$ is everywhere finite, and vanishes sufficiently rapidly,
as $x \rightarrow \infty $, in a sense to be defined below;
(ii)  there are spatial discontinuities demarcating a finite interval;
(iii)  consideration is limited to certain domains of spacetime.
In particular, for propagation from a source point $y$ to an observation
point $x$ in a time $t$, the QNM's give a complete description
for (a) $y$ inside the
interval, $x$ outside the interval, and $t > t_{p}(x,y)$, for a certain
$t_{p}$; or
(b) both $y$ and $x$ inside the interval, and $t > 0 $ (with the retarded
Green's function being zero for $t \le 0 $ ).  Case (b)
is analogous to the completeness of normal modes in hermitian
systems defined on a finite interval.

The above conditions allow dispersion,
backscatter, as well as differences in the damping times, the absence of
which has been conjectured to be important\cite{price}.
The results are valid to all orders in the rate of leakage.
(Lowest order results would be trivial,
since the system becomes hermitian in that limit.)
These results extend our previous work on the wave equation \cite {lly}.

 The spatial coordinate $x$ in (1) often
 represents a radial variable, so first consider a
half-line problem $(x \ge  0 )$ with
$\phi (x=0,t) = 0$. The QNM's are eigen-solutions to the time-independent
KG equation

\begin{equation}
[- \omega ^{2} -  {\partial ^{2}_x} + V(x)] \tilde {\phi} (x) = 0  ~~~~,
\end{equation}

\noindent
with  the boundary conditions $\tilde {\phi} (0) = 0$ and
the outgoing wave condition at infinity.
The eigenfunctions and eigenvalues are denoted as $\tilde {\phi} (x) =
f_{j}(x)$ and $\omega = \omega _{j}$.

The retarded Green's function for the system is defined by  $D G(x,y;t)
= \delta (x - y) \delta (t)$
with $G = 0$ for $t \le  0$,
and the boundary conditions are (i) $G = 0$ for either $x = 0$ or $y = 0$,
and (ii) the outgoing wave condition as either $x \rightarrow \infty $ or
$y \rightarrow \infty $.
In the frequency domain, and henceforth choosing $y \le x$,
$\tilde{G}(x,y;\omega )= f(\omega ,y) g(\omega ,x) / W(\omega )$,
where $f$ and $g$ are solutions to (2) with the boundary conditions
$f(\omega ,x=0) =  0 ; f'(\omega ,x=0) = 1$\cite{origin}, and
$ g(\omega ,x) \rightarrow  \exp (i \omega  x)$ as
$x\rightarrow \infty $, where $'=d/dx$.

The Wronskian $W(\omega ) = g f' - f g'$ is independent of $x$, and its
zeros are the QNM frequencies $\omega_{j}$, with
$f(\omega _{j},x) \propto g(\omega _{j},x) \propto f_{j}(x)$.
For simple zeros \cite{pole}, the residues of $\tilde{G}(x,y;\omega)$
are $K_{j} = f(\omega_{j},y) g(\omega_{j},x) /
[\partial W(\omega =\omega _{j})/\partial \omega ]$.
Multiple zeros can be handled readily. To analyse
the denominator $\partial W/\partial \omega$,
start with the defining equation for  $f(\omega_{j},x)$
and $g(\omega,x)$. The usual manipulations
lead to

\begin{eqnarray}
\nonumber
&&\int_0^X dx \ f(\omega_j,x)g(\omega,x)\\
=&& \left[ g(\omega,x)f'(\omega_j,x)-g'(\omega,x)f(\omega_j,x) \right] _0^X
/ (\omega^2-\omega_{j}^2)   ~,
\end{eqnarray}

\noindent
where the integral is taken along any contour from $x=0$ to $x=X$.
 Use the outgoing wave condition to evaluate $f$ and $g$ at the upper limit,
and take $\omega \rightarrow \omega_{j}$ by l'Hospital's rule.
This then gives

\begin{eqnarray}
\nonumber
&&\ \ \ \int_0^X dx \ f(\omega_j,x)g(\omega_j,x)\\
&&= - \left[ \partial W / \partial \omega + ifg \right]
_{ \omega = \omega_{j} , x = X } / (2 \omega_{j}) ~~~,
\end{eqnarray}

\noindent
for $X \rightarrow \infty$.
Since $f$ and $g$ are proportional at these poles,
a generalized norm of the QNM's can be defined as

\begin{equation}
\ll f_{j}\mid f_{j} \gg \ \ \equiv \lim_{ X \rightarrow \infty}
\int_0^X dx \ f_{j}(x)^2
+i f_{j}(X)^2 / (2 \omega_{j}) ~,
\end{equation}

\noindent with which the residue of $\tilde{G}$ at
$\omega_{j}$ can be expressed as

\begin{equation}
K_{j} = - f_{j}(x) f_{j}(y)
/ \left[ 2\omega _{j} \ll f_{j}\mid f_{j}\gg \right] ~~~.
\end{equation}

\noindent
The generalized norm, introduced in other contexts
\cite{opticqnm,zel,lam}, involves $f^2$ rather than $|f|^2$, and is therefore
complex. Although each term on the right of (5)
does not have a limit as $X \rightarrow \infty$,
the limit exists for the sum.

Next write $G$ in terms of $\tilde{G}$ by a Fourier integral,
and distort the contour
to a large semicircle in the lower half $\omega$ plane.
One then sees that

\begin{equation}
G(x,y;t) = {i\over 2} \sum^{}_{j} {f_{j}(x) f_{j}(y) e^{- i\omega _{j}t}
\over \omega _{j}\ll f_{j} \mid f_{j}\gg } + I_{c} + I_{s}  ~~.
\end{equation}

\noindent
The sum comes from the zeros of $W( \omega)$, $I_{c}$
is the integral along a semicircle at
infinity, and $I_{s}$ comes from any singularities of $f$ and
 $g$ in the lower half $\omega$ plane.
It remains to show that
(i)  $f$ and $g$ are regular in $\omega$, in which case $I_{s}=0$; and
(ii) $I_{c}$ vanishes if there is a discontinuity in $V(x)$ at some $x = a >0.$

The first step is straightforward, by
appealing to well-known results in the quantum theory of
scattering \cite {newton}.
Since $f$ and $g$ satisfy the Schroedinger equation
with energy $\omega ^{2}$, they are analytic functions
of $\omega $ if the potential is bounded and ``has no tail"
\cite{newton}, in the sense that
$\int^{\infty }_{0} dx \ x \mid V(x)\mid \ \ < \infty$, and
$\int^{\infty }_{0} dx \ x \ e^{\alpha x}\mid V(x)\mid \ \ < \infty \ \
  {\rm  for \ any}\  \alpha  > 0  $ .
Our work on potentials that have a tail, e.g., inverse power laws,
will be reported elsewhere.

Secondly, on the large semicircle $\tilde{G}$  can be estimated
by the WKB approximation.  As  $|\omega| \rightarrow \infty $ WKB
approximation is valid
except at points where the potential has discontinuities, which can be
handled by connecting the WKB solutions across them.
Let $\tilde {\phi} (x) \equiv  \exp [i S(x)]$  be  a
solution of (2); then
$S(x) \approx \pm \int k(x) dx $,
where $k(x) = [ \omega ^{2} - V(x) ]^{1/2}$.
$\tilde{G}$ can then be obtained in terms of $\phi (x)$ and $k(x)$.

Consider a potential with a step discontinuity at $x=a$.
The reflection coefficient is
$R = [S'(a^{-}) - S'(a^{+})]/[S'(a^{-})^* + S'(a^{+})] \sim
\omega ^{-2}$ at high frequencies.
For simplicity we give the argument only for Case b: $ 0 < y \le  x < a$.
It is straightforward to show that

\begin{equation}
\tilde{G}(x,y;\omega ) \simeq  \frac{\sin [I(0,y)][e^{-iI(x,a)} +
Re^{iI(x,a)}]}
{ \sqrt{k(x)k(y)} [e^{-iI(0,a)} + Re^{iI(0,a)}]}
\end{equation}

\noindent where $I(u,v) =\int ^{v}_{u} k(x) dx \approx  \omega (v- u)$.
Now on the semicircle $\omega  = \omega _{R} + i\omega _{I} = C e^{i\theta },
\pi  < \theta  <
2\pi $, as $\omega _{I} = C \sin \theta \rightarrow  - \infty $,
both the numerator and the denominator of $\tilde{G}$ are
dominated by the term proportional to $R$, and

\begin{equation}
\tilde{G}(x,y;\omega ) e^{- i\omega t} \simeq e^{-i \omega(t+x-y) } / \omega
{}~~~.
\end{equation}

\noindent As $C \rightarrow  \infty $, this vanishes for
 $t > 0$.  This conclusion remains valid if $V(x)$ has a discontinuity
only in its $p$-th derivative ($p = 0, 1, 2, ...$), since the
reflection coefficient $R$ would vary as $\omega ^{-(p+2)}$\cite {wkb}.
Thus we have proved (apart from a technical
hitch mentioned below) that for a discontinuous potential,
the QNM's are complete for $t > 0$.  For Case a, i.e. $0 < y < a
< x$, a similar estimate gives, in place of (9),
$e^{-i \omega(t-x-y+2a) } / (R \omega)$.
This vanishes and consequently completeness holds only if $t > t_p (x,y) =
\max (x+y-2a,0)$.

For problems on a full line
 $-\infty < x < - \infty$  (as in the case of Schwarzschild hole),
$f$ is still defined as the solution satisfying the
left boundary condition, which is now a unit outgoing wave as
$x \rightarrow - \infty$; all the arguments remain the same.
Generalization to multiple discontinuities is likewise
straightforward.  In this case, completeness in Case b holds in
the interval between the leftmost and the rightmost
discontinuities.

However, there is one technical hitch.  The QNM sum is in fact
a divergent series, but converges to the correct answer when
regularised in a standard way \cite {hardy}.
The simplest regularization is to first invoke symmetry and only
keep modes with Re $\omega_j >0$ (hereafter denoted as $j > 0$)
in the sum in (7); secondly replace all times
$t$ by $t - i\tau$, with $\tau \rightarrow 0^{+}$.  This implies the
use of a regulating factor $I_j(\tau)=\exp(- \omega_j \tau ) $
multiplying the contribution of the $j$-th QNM.
The need for regularization can be seen as follows.  The
proof that the integrand vanishes on the large semicircle
as $C \rightarrow \infty$ is not valid near the
real axis, in particular, where Im $\omega = O(\log C)$.  In this
part, a factor such as exp(Im $\omega \sigma$) ($\sigma$
real and positive, e.g. $\sigma = t -x-y+2a$)
behaves like a power of $C$ rather than exponentially, and it is
necessary to examine the integrand more carefully \cite {jordan}.
If $R \sim \omega ^{-q}$, the powers of $\omega$ in the integrand for $G$
are inadequate to control the contribution
along the semicircle if $q \ge 1$ \cite {convergent}.
For the KG equation
with $V(x)$ discontinuous in its $p$-th derivative, $q = p+2$, so
regularization is always necessary.  For the wave equation with
a discontinuity in the $p$-th derivative of the refractive index
$n(x)$, $q = p$, so regularization is not necessary for
$p = 0$.  The difference by two powers can be understood with
the transformation relating the wave equation to
the KG equation\cite{lly}.

For large $p$, the divergence is more
severe, the regulated sum converges more slowly to the correct
result as $\tau \rightarrow 0$, and the use of a
small finite $\tau$ becomes increasingly
inaccurate.  Thus, in practice, the QNM sum is only useful
for relatively strong discontinuities, e.g. $p = 0, 1$.

To summarize, we have proved, under the conditions stated,
that the Green's function $G$ is expressible, in a certain domain of
$x,$ $y,$ $t$, as the regulated QNM sum

\begin{equation}
G = \lim_{ \tau \rightarrow 0} {\rm Re}
\left[i \sum_{j>0}{f_j(x)f_j(y) e^{-i \omega_{j} t}I_{j}(\tau)
\over \omega_j \ll f_j|f_j \gg }  \right]   ~~.
\end{equation}

\noindent
Considering $\dot{G}$ as $t \rightarrow 0^{+}$,

\begin{eqnarray}
\nonumber
&&\ \ \ \delta (x-y) =\dot{G}(x,y;t=0^+)  \\
&&= \lim_{ \tau \rightarrow 0} {\rm Re} \left[ \sum_{j>0}
{f_j(x)f_j(y)  I_j(\tau) \over \ll f_j|f_j \gg } \right]
\end{eqnarray}

\noindent
valid for $x$, $y$ $\in [0,a]$. This is perhaps the more familiar notion of
completeness
\cite {convergent}.

We now focus on (11), noting that such
a resolution of the identity will be useful for a variety
of problems similar to those of hermitian systems.
The QNM representation of $G$ for $t > 0$
will be further discussed elsewhere in the context of initial value problems.
As an example, consider a potential defined on the full line,
with $V(x) = V_1 > 0 \ \  {\rm for}  \ \ 0 \leq x \leq a $, and zero
otherwise.  This potential has discontinuities and no tail,
thus satisfying the conditions necessary for our results.
The QNM's of this system are easily found.
We consider (11), but as an approximate equality
for some small nonzero regularization parameter $\tau$, and in a distribution
sense, e.g.

\begin{eqnarray}
\nonumber
&&{\rm Re} \sum^{}_{j>0} f_{j}(x) \int^{y_2}_{y_1} dy f_{j}(y) I_j(\tau)
/ \ll f_{j} \mid f_{j}\gg \\
&&\simeq \theta(x-y_1)-\theta(x-y_2)  ~~~,
\end{eqnarray}

\noindent
for $x$, $y_1$, $y_2$ $\in [0,a]$. Figure 1 shows that for small
$\tau$, the partial sum indeed converges close to the
right hand side.  Without the regulator, the sum does not
converge.

This example further illustrates why the series needs to be
regulated. Suppose that for a full-line problem,
the reflection coefficients on both sides behave as
$R(\omega) \sim \omega^{-q}$
at high frequencies ($q=2$ for the present situation).
By setting the denominator in (8) to zero \cite {twoside}, it
is straightforward to show that Im $\omega_j a \sim -q \log |j|$.
In the region $0 < x,y < a$, a typical
term in the product $f_{j}(x)f_{j}(y)$ goes as

\begin{equation}
|f_{j}(x)f_{j}(y)| / \ll f_{j} \mid f_{j} \gg
\sim (|j| \pi)^{q \Lambda/a} ~~~,
\end{equation}

\noindent
where $\Lambda = \max (|x+y-a|,|x-y|)$.
Since the maximum value of $\Lambda$ is $a$,
without the regulator the worst behavior of the summand
in (10) is  $|j|^{q-1}$.
The sum would not converge \cite {convergent}
without a regulating
factor if $q \ge  1$.  The culprit is Im $\omega_{j}a \sim - q \log |j|$,
 from which one
can say unequivocally that the need for regulating the sum is an intrinsic
property of KG open systems.

We have shown that for potentials that have no tail but which
contain spatial discontinuities, the QNM sum is complete,
if suitably regulated.  This result settles a question
in the literature \cite {price}.  The generic
need for regularization, in particular,
distinguishes the KG equation from the wave equation.
When there is a tail, there will be
scattering of the wave from asymptotic regions of space,
leading to power-law type long time behavior,
which will be discussed elsewhere.

In so far as the Green's function $G$ provides the solution to all the
dynamics, the QNM expansion of $G$ will lead to a variety of physical
applications, much as the normal mode expansion of hermitian systems.

We thank RK Chang, HM Lai, KL Liu, SY Liu and RH Price for discussions.
The work of ESCC, PTL and KY is supported in part by a grant from
the Croucher Foundation.  The work of
WMS is supported by the US NFS (Grant no. PHY 94-04788).

\begin{figure}
\caption{Smooth envelope forming upper bound of $\log_{10} E$
 vs $J/10^3$, where $E$ is the difference between the
 right hand side of (12)  and the partial sum
on the left hand side, up to $J$ terms.
  The parameters are $V_1=10^2$, $a=1$,
$y_1=0,$ $y_2=0.5$ and
 $x=0.25,$ $\tau=5 \times 10^{-3}$ (curve 1);
 $x=0.25,$ $\tau=5 \times 10^{-4}$ (curve 2);
 $x=0.75,$ $\tau=5 \times 10^{-3}$ (curve 3);
 $x=0.75,$ $\tau=5 \times 10^{-4}$ (curve 4).}
\end{figure}


\begin{references}
\bibitem{kg} For linearized waves (including KG scalar,
electromagnetic and gravitational waves) in static spherical symmetric
spacetimes, the propagation is exactly described by (1).
\bibitem{lly} Our work on the wave equation
 $ \left[  n^2 (z) {\partial^2 _t}
     -{\partial^2 _z} \right] \psi (z,t) = 0  $
has been reported in Ref. 1.
The wave equation is related to the KG equation (1)
by a transformation  $ dx/dz = n(z), \psi  = n^{-1/2} \phi $ ,
with the potential related to $n(z) $ by
$V = (2n^3)^{-1}(d^2 n/dz^2) -(3/4n^4)(dn/dz)^2 \ . $  However, there
are two important differences : (a) A continuous
$V(x)$ implies a continuous $n(z)$ but not the other way around, (b) the need
to regulate the QNM sum in the KG case is generic, see below.
\bibitem{origin} We assume that here $f'$ is nonzero at the origin. Otherwise,
if $f \sim x^{l+1} $ as $x \rightarrow 0$ then we require $\lim_{x \rightarrow
0}
x^{-(l+1)}f(\omega,x) = 1$.
\bibitem{wkb} If there are discontinuities of any order, a connection
formula has to be used, as the WKB series is not valid at those points.
\bibitem{jordan} Note that Jordan's lemma states that an integral
$\int d\omega  e^{- i\omega t} f(\omega )$ along a
large semicircle in the lower half plane vanishes for $t > 0,$ provided that
$\left|f(\omega )\right| \rightarrow 0$ at infinity.  The necessary bound on
$f$ arises similarly from
the need to bound the part with
$\mid \omega _{I}\mid  = O(\log \omega _{R}).$
\bibitem{convergent} To be specific, in this Letter we consider
pointwise convergence for $G$ (which is a well defined
quantity) and convergence in the distribution sense for the
representation of $\delta (x-y)$, which comes from $\dot G$.
Going from $G$ to $\dot G$ costs a factor $\omega$, but going
from the pointwise to the distribution
sense effectively gains a factor $\omega ^{-1}$.  This happens
because asymptotically the typical dependence is $e ^{i \omega y}$,
so integration over $y$ gives a factor $\omega ^{-1}$.  Thus
the pointwise validity of (10) and the validity of (11) in a
distribution sense require the same conditions.
\bibitem{twoside}  Note that (8) has to be slightly modified for
this full line case.
\end{references}
\end{document}